   \documentclass[prl,aps,twocolumn,showpacs]{revtex4}
\usepackage[dvips]{graphicx}
\usepackage{dcolumn}
\newcommand{\be}{\begin{equation}}
\newcommand{\ee}{\end{equation}}
\newcommand{\bea}{\begin{eqnarray}}
\newcommand{\eea}{\end{eqnarray}}
\newcommand{\nn}{ \nonumber}
\newcommand{\ds}{\displaystyle}
\begin{document}

\title{Fermi-liquid  and Fermi surface geometry effects in propagation of 
low frequency electromagnetic waves through thin metal films }

\author{ Natalya A. Zimbovskaya}
\affiliation{Department of Physics and  Electronics, University of Puerto 
Rico, Humacao, PR 00791}

\begin{abstract}
In the present work we theoretically analyze the 
contribution from a transverse Fermi-liquid collective mode to the 
transmission of electromagnetic waves through a thin film of a clean 
metal in the presence of a strong external magnetic field. We show that 
at the appropriate Fermi surface geometry the transverse Fermi-liquid 
wave may appear in conduction electrons liquid at frequencies $ \omega $ 
significantly smaller than the cyclotron frequency of charge carriers 
$ \Omega $ provided that the mean collision frequency $ \tau^{-1}$ is 
smaller than $\omega. $ Also, we show that  in realistic metals size 
oscillations in the transmission coefficient associated with the 
Fermi-liquid mode may be observable in experiments. Under certain 
conditions these oscillations may predominate over the remaining size 
effects in the transmission coefficient.
\end{abstract}

\date{\today}
\maketitle

\section{i. introduction}

It is well known that electromagnetic waves incident on the surface of 
a metal cannot penetrate inside the metal deeper than a thin surface 
layer (skin layer). This happens due to the damping effect of conduction 
electrons absorbing the wave energy via dissipationless Landau damping 
mechanism \cite{1}. A strong magnetic field ${\bf B} = (0,0,B)$ applied 
to the metal restricts the motion of electrons in the $x,y$ plane, 
creating ``windows of transparency". These windows are regions in the 
$q,\omega$ plane $(q,\omega$ are the wave vector and the frequency of 
the electromagnetic wave, respectively) where the Landay damping cannot 
take place. As a result, in the presence of the external magnetic field 
various weakly attenuating electromagnetic waves, such as helicoidal, 
cyclotron and magnetohydrodynamic waves may propagate in the electron 
liquid of a metal \cite{2,3}. 

Fermi-liquid (FL) correlations of 
conduction electrons bring changes in the wave spectra. Also, new 
collective modes may appear in metals due to FL interactions among 
the electrons. These modes solely occur owing to the FL interactions, 
so they are absent in a gas of charge carriers. Among these modes there 
is the Fermi-liquid cyclotron wave first predicted by Silin \cite{4} and observed in alkali metals \cite{5,6}. 
In a metal with the nearly spherical Fermi surface (FS) this mode is the 
transverse circularly polarized wave propagating along the external 
magnetic field whose dispersion within the collisionless limit 
$(\tau \to \infty)$ has the form \cite{7}:  
\be 
\frac{\omega}{\omega_0}=1 + \frac{8}{35}\frac{1}{\alpha} (qR)^2
   \ee
  Here, $ R=v_0/\Omega;\ v_0 $ is the maximum value of the electron 
velocity component along the magnetic field (for the spherical FS 
$v_0 $ equals to the Fermi velocity $v_F);\ \Omega = e B/mc $ is the cyclotron 
frequency, $\tau $ is the electron scattering time, and the dimensionless 
parameter $\alpha $ characterizes FL interactions of conduction electrons.
For  the spherical FS the electrons cyclotron mass coincides with their effective mass $m.$ The difference between the frequency $\omega_0 =\omega(0) $ and the cyclotron frequency is determined with the value of the Fermi-liquid 
parameter $\alpha,$ namely: $\omega_0 =\Omega(1 + \alpha).$ Depending 
on whether $\alpha $ takes on a positive/negative value $\omega_0$ is 
greater/smaller than $\Omega. $ Further we assume for certainty that 
$\alpha < 0. $ When 
$qR \ll 1$ the dispersion curve of this Fermi-liquid 
cyclotron wave is situated in the window of transparency whose boundary 
is given by the relation: $\omega = \Omega - qv_0$ which corresponds to 
the Doppler-shifted cyclotron resonance for the conduction electrons. 

\begin{figure}[t]
\begin{center}
\includegraphics[width=8.8cm,height=4.7cm]{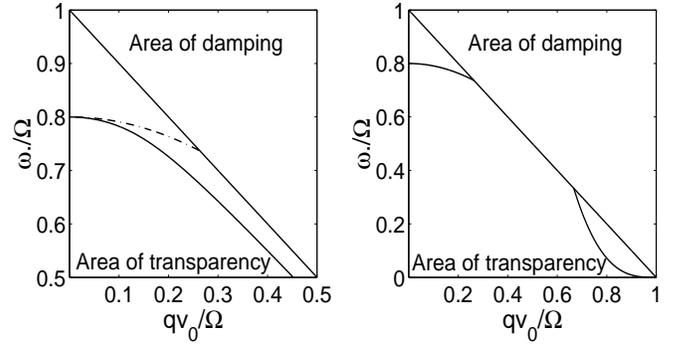}
\caption{Left panel: Dispersion of the transverse Fermi-liquid cyclotron 
wave traveling along the external magnetic field for the spherical 
(dash-dot line) an paraboloidal (solid line) FSs. The curves are plotted 
using Eqs. (1),(3) assuming $ \alpha = -0.2. $ Right panel: A schematic 
plot of the dispersion of a transverse Fermi-liquid mode in a metal whose 
FS includes nearly paraboloidal segments. The low frequency 
$(\omega \ll \Omega) $ branch is shown along with the cyclotron wave. 
For both panels the straight line  corresponds to the Doppler-shifted 
cyclotron resonance.}
 \label{rateI}
\end{center}
\end{figure}

 This is shown in the Fig. 1 (left panel). However, the dispersion curve 
meets the boundary of the transparency region at $ q=q_m \approx 
5|\alpha|/3R$ \cite{7}, and at this value of $q$ the dispersion curve 
is terminated \cite{8}. So, for reasonably weak FL interactions 
$ |\alpha| \sim 0.1 $ the Fermi-liquid cyclotron wave may appear only at 
$qR \ll 1 $ and its frequency remains close to the cyclotron frequency 
for the whole spectrum \cite{9}. Similar conclusions were made using some other 
models to mimic the FS shape such as an ellipsoid, a nearly ellipsoidal 
surface and a lens made out of two spherical segments \cite{10,11}.

It is clear that the main contribution to the formation of a weakly 
attenuated collective mode near the boundary of the transparency region 
at $\omega \ll \Omega $ comes from those electrons which move with the 
greatest possible speed along the magnetic field $\bf B.$ The greater 
is the relative number of such electrons the more favorable conditions 
are developing for the wave to emerge and to exist at comparatively low 
frequencies $ \tau^{-1} \ll \omega \ll \Omega. $ The relative number of 
such ``efficient" electrons is determined with the FS shape, and the best 
conditions are reached when the FS includes a lens made out of two 
paraboloidal cups. Such lens corresponds to the following energy-momentum 
relation for the relevant conduction electrons:
  \be 
  E{\bf (p)} = \frac{{\bf p}^2_\perp}{2m_\perp} + v_0|p_z|  
\ee
  where $ p_z, \bf p_\perp $ are the electron quasimomentum components in 
the plane perpendicular to the external magnetic field $ {\bf B} = (0,0,B), $ 
and along the magnetic field, respectively. The effective mass $m_\perp$ corresponds to electrons motions in the $x y$ plane.
  This model was employed in some earlier works to study transverse 
collective modes occuring in a gas of charge carriers near the 
Doppler-shifted cyclotron resonance which are known as dopplerons  
\cite{12,13,14}. It was shown \cite{15} that for negative values of the 
Fermi-liquid parameter $ \alpha, $ and provided that the FS contains a 
paraboloidal segment described by the Eq. (2) the dispersion of the 
transverse Fermi-liquid wave propagating along the magnetic field has 
the form $(\tau \to \infty):$
   \bea 
\frac{\omega}{\Omega}&=& 1 -\frac{1}{2}(qR + |\alpha|)
 \nn\\ & - &
\frac{1}{2}\sqrt{(qR -|\alpha|)^2 + \frac{4}{3} \frac{|\alpha|(qR)^2}{qR 
+ \sqrt{(qR)^2 + |\alpha|^2}}} .
  \eea 
 where $\Omega =eB/m_\perp c. $
  This result shows that for the paraboloidal FS there are no limitations 
on frequency of the Fermi-liquid cyclotron wave within the collisionless 
limit (see Fig. 1, left panel). The only restriction on the wave frequency 
is caused by the increase  of the wave attenuation due to collisions.   
Taking into account electron scattering
one can prove that the wave is 
weakly attenuated up to a magnitude of the wave vector of the order of 
$ \Omega (1 - 1 / |\alpha|\Omega \tau)/ { v_0}. $ This value (especially 
for small $ |\alpha|) $ is significantly larger than the value $ q _ m $ 
for the spherical Fermi surface.  Therefore,  the frequency of the 
Fermi-liquid cyclotron waves for negative $ \alpha $ can be much smaller 
than $ \Omega $ (remaining greater than $ 1/\tau $). Comparing the 
dispersion curves of the transverse Fermi-liquid cyclotron wave for 
spherical and paraboloidal FSs we see that the FS geometry strongly 
affects the wave spectrum, and it may provide a weak attenuation of 
this mode at moderately low frequencies $ \omega \ll \Omega. $ In the 
present work we concentrate on the analysis of the effects of the FS 
geometry on the occurence of weakly damped Fermi-liquid cyclotron waves 
propagating in metals along the applied magnetic field at low frequencies 
$(\tau^{-1} \ll \omega \ll \Omega).$

 We show below that in realistic 
metals with appropriate FSs one may expect a low frequency Fermi-liquid 
mode to occur along with the Fermi-liquid cyclotron wave as presented in 
the Fig. 1 (right panel). Both waves have the same polarization, and 
travel in the same direction. Also, we consider possible manifestations 
of these low frequency Fermi-liquid waves  estimating the magnitude of 
the corresponding size oscillations in the transmission coeficient for 
electromagnetic waves propagating through a thin metal film.

\section{ii. dispersion equation for the transverse Fermi-liquid waves}

 In the following analysis we
restrict our consideration with the case 
of an axially symmetric Fermi surface whose symmetry axis is parallel to 
the magnetic field. Then the  response of the electron liquid of the metal 
to an electromagnetic disturbance  could be expressed in terms of the 
electron conductivity circular components $ \sigma_\pm (\omega,{\bf q}) = 
\sigma_{xx} (\omega,{\bf q} )\pm i \sigma_{yx} (\omega,{\bf q} ). $ The 
above restriction on the FS shape enables us to analytically calculate 
the conductivity components.
 Also, the recent analysis carried out in 
Ref. \cite{16} showed that no qualitative difference was revealed in the 
expressions for the principal terms of the surface impedance computed for 
the axially symmetric FSs and those not possessing such symmetry, provided 
that $\bf B $ is directed along a high order symmetry axis of the Fermi 
surface.  This gives grounds to expect the currently employed model to 
catch main features in the electronic response which remain exhibited 
when the FSs of generalized (non axially symmetric) shape are taken into 
consideration.

Within the phenomenological Fermi-liquid theory electron-electron 
interactions are represented by a self-consistent field affecting any 
single electron included in the electron liquid. Due to this field the 
electron energies $ E\bf (p) $ get renormalized, and the renormalization 
corrections depend on the electron position $ \bf r $ and  time $ t: $
   \be 
 \Delta E = T r_{\sigma'} \int \frac{d^3 \bf p'}{(2 \pi \hbar)^3}\, 
F ({\bf p, \hat \sigma; p', \hat \sigma') \delta \rho (p', r,\hat\sigma',}
 t) .
  \ee
  Here, $ \delta \rho {\bf (p, r, \hat\sigma,} t) $ is the electron 
density matrix, $ \bf p $ is the electron quasimomentum, and $ \hat\sigma$ 
is the spin Pauli matrix. The trace is taken over spin numbers $\sigma. $ 
The Fermi-liquid kernel included in Eq. (4) is known to have a form:   
\be 
 F ({\bf p, \hat\sigma; p', \hat\sigma') = \varphi (p,p')} + 
4 \bf (\hat\sigma \hat\sigma') \psi (p,p')
  \ee 
   For an axially symmetric FS the functions $ \varphi \bf (p,p')$ and 
$ \psi \bf (p,p')$  do not  vary under identical change in the directions of 
projections $ \bf p_\perp $ and $ \bf p_\perp' .$  These functions 
actually depend only on cosine of an angle $\theta$ between the vectors 
$\bf p_\perp $ and $\bf p_\perp'$ and on the longitudinal components of 
the quasimomenta $ p_z $ and $ p_z'$.

We  can separate out even and odd 
in  $\cos \theta $ parts of the Fermi-liquid functions. Then the function 
$ \varphi \bf (p,p')$ can be presented as follows:
   \be
 \varphi({\bf p,p'}) = \varphi_0 (p_z,p_z', \cos\theta) + 
({\bf p_\perp p_\perp'}) \varphi_1 (p_z,p_z',\cos\theta), 
  \ee 
  where $ \varphi_0,\varphi_1 $ are even functions of $ \cos\theta.$ 
Due to invariancy of the FS under  the replacement 
$\bf p \to - p $ 
and $\bf p'\to -p',$ the functions $ \varphi_0 $ and $ \varphi_1 $  
should not vary under simultaneous change in signs of $ p_z $ and $ p_z'.$ 
Using this, we can subdivide the functions $ \varphi_0,\varphi_1 $ into 
the parts which are even and odd in $ p_z, p_z',$ and to rewrite Eq. (6) 
as:
   \be
 \varphi (p_z,p_z',\cos \theta) = \varphi_{00} + p_z p_z'\varphi_{01} 
+ ({\bf p_\perp p_\perp'}) (\varphi_{10} + p_z p_z'\varphi_{11}). 
  \ee
  The function $ \psi \bf (p,p')$ may be presented in the similar way. 
In the Eq. (7) the functions $ \varphi_{00},\varphi_{01}, \varphi_{10},
\varphi_{11} $ are even in all their arguments, 
namely: $ p_z,p_z'$ and 
$ \cos \theta .$ 

In the following computation of the electron 
conductivity we employ the linearized transport equation for the 
nonequilibrium distribution function $ g {\bf (p,r,} t) = Tr_\sigma 
(\delta \rho {\bf ( p, r,\hat\sigma,} t)). $ While considering a simple 
harmonic disturbance $ {\bf E = E}_{q\omega}\exp (i {\bf q\cdot r}- i 
\omega t), $ we may represent the coordinate and space dependencies of 
the distribution function $ g {\bf (p,r,}t ) $ as
 $ g {\bf (p,r,}t) = g_{q\omega} \exp (i {\bf q r}- i\omega t). $ Then the 
linearized transport equation for the amplitude $ g_{q\omega} \bf (p) $ 
takes on the form:
  \be 
 \frac{\partial g^e_{q\omega}}{\partial \tilde t} + i{\bf q \cdot v} 
g_{q\omega}^e + \Big(\frac{1}{\tau} -i\omega\Big)g_{q\omega} + 
e\frac{\partial f_{\bf p}}{\partial E_{\bf p}} {\bf v E}_{q\omega} = 0. 
  \ee
   Here, $f_{\bf p} $ is the Fermi distribution function for the electrons 
with energies $ E\bf (p), $ and $ {\bf v} = \partial E/\partial \bf p $ is 
the electrons velocity. The collision term in the Eq. (8) is written using 
the $ \tau $ approximation  which is acceptable for high frequency 
disturbances $(\omega \tau \gg 1) $ considered in the present work. The 
derivative $ \partial g_{q\omega}^e/\partial \tilde t $ is to be taken 
over the variable $ \tilde t $ which has the meaning of time of the 
electron motion along the cyclotron orbit. The function $ g_{q\omega}^e 
\bf (p)$ introduced in the Eq. (8) is related to $ g_{q\omega} \bf (p)$  
as follows:
  \be 
   g_{q\omega}^e {\bf (p)} = g_{q\omega} {\bf (p)} 
- \frac{\partial f_{\bf p}}{\partial E_{\bf p}} \sum_{\bf p'} \varphi 
{\bf (p,p')} g_{q\omega} {\bf (p')} .
  \ee
  So, the difference between 
the distribution functions $g_{q\omega} {\bf (p)} $ and $g_{q\omega}^e 
{\bf (p)} $ originates from the FL interactions in the system of 
conduction electrons.

Using the transport equation (8) one may derive 
the expressions for $ \sigma_\pm (\omega,\bf q) $ including terms 
originating from the Fermi-liquid interactions. The computational details 
are given in the Refs. \cite{17,18}. The results for the circular 
components of the conductivity for a singly connected FS could be written 
as follows:
   \be 
  \begin{array}{ll} 
 \sigma_\pm = &\ds \frac{2ie^2 A(0)}{(2\pi\hbar)^3 q} \\\\
  &  \ds \times
 \frac{\left[\ds \Phi_0^\pm \Big(1- \frac{\alpha_2u}{Q_2} \Phi_2^\pm \Big) 
+\frac{\alpha_2 u}{Q_2} (\Phi_1^\pm)^2 \right]}{\left[\ds \Big(1- 
\frac{\alpha_1 u}{Q_0} \Phi_0^\pm \Big) 
 \Big(1- \frac{\alpha_2 u}{Q_2} \Phi_2^\pm\Big) 
 +\frac{\alpha_1\alpha_2}{Q_0 Q_2} u^2 (\Phi_1^\pm)^2 \right]}. 
  \end{array}   \ee
  Here,
  \bea 
   \Phi_n^\pm = \int_{-1}^1 \frac{\overline a(x) \overline m_\perp (x) 
x^n dx }{u\chi_\pm \mp \overline v(x)},
  \\ \nn\\  
Q_n = \int_{-1}^1 \overline a(x) \overline m_\perp (x) x^n dx .
    \eea
   \be 
   \begin{array}{ll} 
 \overline a(x) ={A(x)}/{A(0)}, &\quad \overline v(x)={v_z}/{v_0},
   \\
\overline m_\perp (x) = {m_\perp (x)}/{m_\perp (0)}, & \quad x = {p_z}/{p_0},
  \\
 \chi_\pm = 1 \pm {\Omega}/{\omega} + {i}/{\omega\tau},  & \quad u ={\omega}/{q v_0}
   \end{array}
  \ee
  where $ v_0,\ p_0 $ are the maximum values of longitudinal components of the electron quasimomentum and velocity; $ A(x) $ is the FS cross-sectional area; $ m_\perp (x) $ is the cyclotron mass of electrons. The dimensionless factores $ \alpha_{1,2} $ in the Eq. (10) are related to the Fermi-liquid parameters $ \varphi_{10} $ and $ \varphi_{11}:$
   \be 
  \alpha_{1,2} = f_{1,2} \big/(1 + f_{1,2})
 \ee
   where
  \bea 
 &&f_1 =\frac{2}{(2\pi \hbar)^3} \int p_\perp^2 \varphi_{10} m_\perp dp_z, 
  \nn \\
  && f_2 = \frac{2}{(2\pi \hbar)^3} \int p_\perp^2 p_z^2\varphi_{11} m_\perp dp_z.
     \eea

When an external magnetic field is applied, electromagnetic waves may travel inside the metal. In the present work we are interested in the transverse waves propagating along the magnetic field. The corresponding dispersion equation has the form:
  \be
  c^2 q^2 - 4 \pi i \omega \sigma_\pm (\omega,{\bf q} ) = 0 .
  \ee
  When dealing with the electron Fermi-liquid, this equation for $``-"$ polarization  has  solutions corresponding to helicoidal waves and the transverse Fermi-liquid waves traveling along the magnetic field. While the relevant charge carriers are holes the $``+" $ polarization is to be chosen in the Eq. (16).

Considering these waves we may simplify the dispersion equation (16) by omitting the first term. Also, we can neglect corrections of the order of $ c^2 q^2/\omega_p^2 \ (\omega_p $ is the electron plasma frequency) in the expression for the conductivity. Then the Fermi-liquid parameter $ \alpha_1 $ falls out from the dispersion equation, and the latter takes on the form:
   \be 
  \Delta (u) = 1/\alpha_2
   \ee
  where $ \Delta (u) = \ds \frac{u}{Q_2} \big[\Phi_2^- -(\Phi_1^-)^2/\Phi_0^-\big] $.

 Assuming the mass $m_\perp $ to be the same over the whole FS, and
 expanding the integrals $ \Phi_n^- $ in powers of $ u^{-1} $ and keeping terms of the order of $ u^{-2} $ we get the dispersion relation for the cyclotron mode at small $q\ (u\gg 1):$
  \be 
  \omega = \Omega (1 + f_2) \bigg[1 + \frac{\eta}{f_2} \bigg(\frac{q v_0}{\Omega} \bigg)^2 \bigg].
  \ee
 where:
  \be 
 \eta = \left[\int_{-1}^1 \overline a(x) \overline v^2 (x) x^2 dx - \frac{1}{Q_0} \left(\int_{-1}^1 \overline a (x)\overline v(x) x dx \right)^2 \right]\frac{1}{Q_2}.
 \ee
 For an isotropic electron liquid $ \eta = 8/35, $ and the expression (18) coincides with the expression (1) where $\alpha = f_2.$ 
 Also, adopting the model (2) we may analytically calculate the integrals $ \Phi_n^- $ and to transform the dispersion equation (17) as:
  \be 
  3(\chi_- + \alpha_2)(1 - (u \chi_-)^2) = \alpha_2.
  \ee
 At small negative values of the parameter $ \alpha_2$ this equation has a solution of the form (3) where $ \alpha = \alpha_2. $

Now, we start to analyze possibilities for the low frequency $(\tau^{-1} \ll \omega \ll \Omega)$ transverse Fermi-liquid mode to emerge in realistic metals where the cyclotron mass depends on $p_z.$ Such waves could appear near the Doppler-shifted cyclotron resonance. Assuming $ \alpha_2 < 0 $ we may describe the relevant boundary of the transparency region by the equations:
   \be 
          \left \{  \begin{array}{l} 
S(\omega, q , p_z) = 0, \\\\
{\displaystyle 
{\partial S (\omega, q, p_z)}/{\partial p_z} = 0,}
\end{array} \right.
                     \ee
 where $ S(\omega, q, p_z) = \omega - \Omega (p_z) + q v_z (p_z). $ For small $ \omega $ we have
         \be 
\left \{  \begin{array}{l} 
{\displaystyle \Omega (p_z) 
\left (1 + \frac{c q}{2 \pi |e| B} \frac{d A}{d p_z} \right)
= 0,} \\\\
{\displaystyle 
\frac{d \Omega}{d p_z}
\left (1 + \frac{c q}{2 \pi |e| B} \frac{d A}{d p_z} \right)
+  \frac{\Omega (p_z)c q}{2 \pi |e| B} \frac{d^2 A}{d p_z^2} = 0.}
\end{array} \right.
                 \ee
 We see that the attenuation at the boundary for small $ \omega $ is carried out by the electrons belonging to neighbourhoods of particular cross-sections on the Fermi surface where extrema of the value $ d A/d p_z $ are reached. These can be neighbourhoods of limiting points or lines of inflection, as shown in the figure 2.

\begin{figure}[t]
\begin{center}
\includegraphics[width=8.8cm,height=4.7cm]{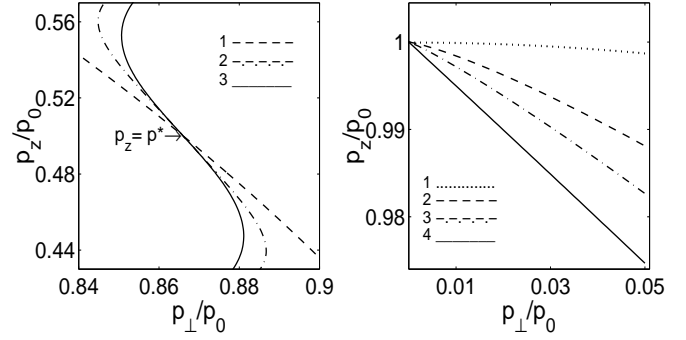}
\caption{ Schematic plots of the FS profiles in the vicinities of inflection lines (left panel) and vertices (right panel). Left panel: The profiles are drawn in accordance with the Eq. (23) assuming $ p^* = 0.5 p_0, \ |d^s \overline a /dx^s|_{x=x^*} =
 |d \overline a /dx|_{x=x^*} =1 ,$  and $ s=5 $ (curve 2), $ s=3 $ (curve 3). The curve 1 corresponds to a paraboloidal strip on the FS near $ x^* =0.5 \ (s\to\infty). $ Right panel: The curves are plotted asuming $x^* =1,\ \overline a (1) =0. $ The curves 1 and 4 correspond to a spherical and paraboloidal FSs, respectively; the curves 2,3 represent nearly paraboloidal FSs with $ s=7,9 $, respectively.  }
 \label{rateI}
\end{center}
\end{figure}

 In general, to study various effects in the response of electron liquid of metal near the Doppler-shifted cyclotron resonance  one must take into account contributions from all segments of the FS, therefore the expressions for the conductivity components (10) are to be correspondingly generalized. However, in studies of our problem it is possible to separate out that particular segment of the FS where the electrons producing the low frequency Fermi-liquid wave belong. The contribution from the rest of the FS is small, and we can omit it, as shown in Ref. \cite{15}.

So, in the following studies we may use the dispersion equation (17) where the integrals $ \Phi_n^\pm $ are calculated for the appropriate segment of the FS.
 It follows from this equation  that the dispersion curve of
the cyclotron wave will not intersect the boundary of the region of transparency when the function $ \Delta (u) $ diverges there. A similar analysis was carried out in the theory of dopplerons  \cite{14}.  It was proven that when the appropriate component of the conductivity (integral of a type of $ \Phi _ 0 (u)) $ goes to infinity at the Doppler-shifted cyclotron resonance, it provides the propagation of the doppleron without damping in
a broad frequency range.

 In the further analysis we assume for certainty that the extrema of $ d A/d p_z $ are reached at the inflection lines $ p_z=\pm p^*. $ In the vicinities of these lines we can use the following approximation:
  \be 
  \overline a(x) \approx \overline a(x^*) + \frac{d\overline a}{dx}\bigg|_{x=x^*} \!\!({x\mp x^*}) \pm \frac{1}{s!}\frac{d^s \overline a}{d x^s} \bigg |_{x=x^*} \!\!({x \mp x^*})^s .
   \ee
  In this expression $ x^* = p_z/p^*, $
   and the parameter $ s \ (s \geq 3) $ characterizes the FS shape near the inflection lines at $ x = \pm x^*. $ The greater is the value of $ s $ the closer is the FS near $ p_z = \pm p^* $ to a paraboloid (see Fig. 2). When $ s = 1 $ the FS has spherical/ellipsoidal shape in the vicinities of these points.

\begin{figure}[t]
\begin{center}
\includegraphics[width=8.8cm,height=4.7cm]{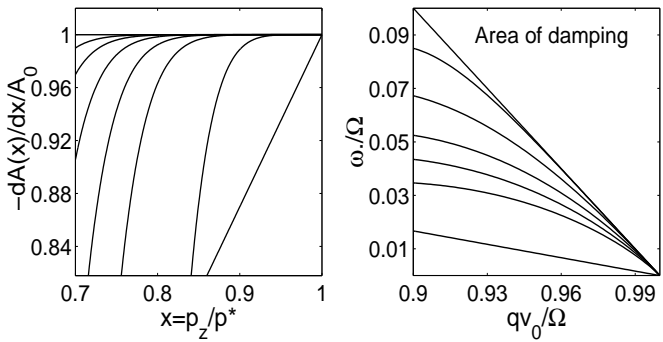}
\caption{Left panel: Dependencies of $ d\overline a/dx $ of $x$ near the inflection line on the FS at $ x=x^* .$ The  curves are plotted for $ s = 4,5,6,7,8,9 $ (from the right to the left).  
 Right panel: Dispersion curves of the low frequency transverse Fermi-liquid waves. The curves are plotted at $ \alpha_2 = -0.2;\ s =4,5,6,7,8,$ and $s\to \infty $ (from the top to the bottom) in the collisionless limit assuming that  $|d^s \overline a / d\overline x^s|_{x=x^*} = |d \overline a / d\overline x|_{x=x^*} = 1$. }
 \label{rateI}
\end{center}
\end{figure}

The dependencies of the derivative $ d\overline a/dx $ of $ x $ near $ x = x^*$ are presented in the left panel of the Fig. 3. In this figure the horizontal line corresponds to a paraboloidal FS $(s \to \infty),$ the straight line on the right is associated with a spherical FS $(s = 1) $, and the remaining curves are plotted for $ (s > 3).$ We can see that the greater is the shape parameter $ s $ the broader are nearly paraboloidal strips in the vicinities of the FS inflection lines. Consequently, the greater number of conduction electrons is associated with the nearly paraboloidal parts of the FS, and this creates more favorable conditions for the wave to occur. Similar analysis may be carried out for the case when $ dA/dp_z $ reachs its extremal values at the vertices  of the FS. Again, to provide the emergence of the transverse low frequency Fermi-liquid mode the FS near $p_z = \pm p_0 $ must be nearly paraboloidal in shape.

Using the asymptotic expression (23) we may calculate the main term in the function $ \Delta (u).$ This term diverges at the boundary of the region of transparency when $ s\geq 3, $ and it has the form:
   \be 
  \Delta_-(u) = - \nu_s u (1 - u \chi_-)^{\mu_s}
  \ee
  where $ \mu_s = (5 - 2s)/(2s -2).$ For $s > 3, \ \mu_s$ takes on negative values, so within the collisionless limit $(\tau\to \infty) $ the function $ \Delta_-(u) $ diverges when $ 1- u\chi_- \to 0. $ The value of the factor $ \nu_s $ is determined with the FS geometry near the inflection line, namely:
  \be 
 \nu_s = \frac{\pi \overline a (x^*) \overline m_\perp (x^*)\zeta_s}{Q_2(s-1) \sin [3\pi/(2s-2)]}
  \ee
 where
  \be
  \zeta_s = \left(\frac{|d^s\overline a/d x^s|_{x=x^*}}{(s-1)! |d\overline a/dx|_{x=x^*}}\right)^{-3/2(s-1)}.
  \ee

Now, we can employ the approximation (24) to solve the dispersion equation (17). The solutions of this equation within the collisionless limit describing the low frequency transverse Fermi-liquid wave at different values of the shape parameter $ s $ are plotted in the figure 3 (right panel). All dispersion curves are located in between the boundary of the transparency window and the straight line corresponding to the limit $ s\to \infty $ (a paraboloidal FS). The greater is the value of $ s $ the closer is the dispersion curve to this line.

So, we showed that the low frequency $(\omega \ll \Omega)$ transverse Fermi-liquid wave could appear in a metal put into a strong $ (\Omega \tau \gg 1)$ magnetic field. This could happen
 when the FS is close to a paraboloid near those cross-sections where $dA/dp_z $ reachs its maxima/minima. Therefore, the possibility for this wave to propagate in a metal is provided with the local geometry of the Fermi surface near its inflection lines or vertices. 

When $\Omega $ depends on $ p_z $ and $ \omega $ increases, electrons associated with various cross-sections of the Fermi surface participate in the formation of the  wave.   To provide the divergence of the function $ \Delta (u) $ near the Doppler-shifted cyclotron resonance we have to require that not merely narrow strips near lines of inflection or vicinities of limiting points but rather large segments of the Fermi surface are nearly paraboloidal. This condition is too stringent for FSs of real metals.  So, we can expect that the dispersion curve of the low frequency transverse Fermi-liquid wave  intersects the
boundary of the region of transparency at rather small $ \omega ,$
as shown in the right panel of the Fig. 1.

\section{iii. size oscillations in the surface impedance}

 To clarify possible manifestations of the considered Fermi-liquid wave in experiments we calculate the contribution of these waves to the  transmission coefficient of a metal film.
We assume that  the film occupies the region $ 0\ll z \leq L $ in the presence of an applied magnetic field directed along a normal to the interfaces. An incident electromagnetic wave with the electric and magnetic components $ {\bf E} (z) $ and ${\bf b} (z) $ propagates along the normal to the film. Also, we assume that the simmetry axis of the FS is parallel to the magnetic field $ (z$-axis) and the interfaces reflect the conduction electrons in a similar manner. Then the Maxwell equations inside the metal are reduced to the couple of independent equations for circular components of the electrical field $ E_\pm (z)  \exp (-i \omega t)$ where $ (E_\pm = E_x \pm i E_y):$
   \bea 
 && \frac{\partial^2 E_\pm}{\partial z^2} =- \frac{\omega^2}{c^2} E_\pm (z) - \frac{4\pi i \omega}{c^2} j_\pm (z),
  \\ \nn\\ &&
   \frac{\partial E_\pm (z)}{\partial z} = \mp \frac{\omega}{c} b_\pm (z).
  \eea   
 Here, $ b_\pm (z) $ and  $ j_\pm (z) $ are the magnitudes of the magnetic component of the incident electromagnetic wave and the electric current density inside the film, respectively. Expanding the magnitudes $ E_\pm (z) $ and $ j_\pm (z)$ in Fourier series we arrive at the following equation for the Fourier transforms:
  \be 
  -\frac{c^2 q_n^2}{4\pi i\omega} E_n^\pm + j_n^\pm =
\mp \frac{ic}{4\pi}[(-1)^n b_\pm (L) - b_\pm (0)]
    \ee
   where $ E_n^\pm $ equals:
   \be 
  E_n^\pm = \int_0^L E_\pm(z) \cos (q_n z) dz
  \ee
 and $ q_n = \pi n/L.$

It was mentioned above that possible frequencies of the low frequency Fermi-liquid mode have to satisfy  the inequality $ (|\alpha_2| \tau)^{-1} \ll \omega < \Omega. $ For $ \tau < 10^{-9} $s the frequency $ \omega $ can not be lower than $10^9\div 10^{10} $s$^{-1}.$
  Due to high density of conduction electrons in good metals the skin depth $ \delta $ may be very small. Assuming the electron density to be of the order of $10^{30}$m$^{-3}$, and the mean free path $l \sim 10^{-5} $m (a clean metal), we estimate the skin depth at the disturbance frequency $ \omega \sim 10^9 $s$^{-1} $ as $ \delta \sim 10^{-6} $m.  Therefore, at high frequencies $ \omega $ the skin effect in good metals becomes extremely anomalous so that $ \delta/l \sim 10^{-1} \div 10^{-2} $ or even smaller. 
  Correspondingly, the anomaly parameter $ \xi = l/\omega\tau \delta $ is of the order $ 10^2\div 10^3 .$ Thus, for all frequency range of the considered Fermi-liquid mode the skin effect is of anomalous character.
  Under these conditions electrons must move nearly in parallel with the metal surface to remain in the skin layer for a sufficiently long while. The effect of the surface roughness on such electrons is rather small. 
Nevertheless, we may expect the effects of surface roughness to bring changes
in the corresponding size oscillations of the transmission coefficient. To take into account the effects of diffuse scattering of electrons from the surfaces of the film one must start from the following expression for the Fourier transforms of the current density components:
   \be 
j_n^\pm = \sigma_n^\pm E_n^\pm + \sum_{n'=0} \left(1-\frac{1}{2} \delta_{n'0} \right) \sigma_{nn'}^\pm E_{n'}^\pm
  \ee
 where $\sigma_n^\pm = \sigma_{xx} (\omega,q_n)\pm i\sigma_{yx} (\omega,q_n) $ are the circular components of the bulk conductivity, and $\sigma_{nn'} = \sigma_{xx} (\omega,q_n,q_{n'}) \pm \sigma_{yx} (\omega,q_n,q_{n'})$ are the circular components of the surface conductivity. The effects originating from the surface roughness are included in $\sigma_{nn'}^\pm $ which becomes zero for a smooth surface providing the specular reflection of electrons. The calculation of $\sigma_{nn'}^\pm$ is a very difficult task which could hardly be carried out analytically if one takes into account Fermi-liquid correlations of electrons. However, such calculations were performed for a special case of paraboloidal  FS corresponding to the energy-momentum relation (2) in the earlier work \cite{19}. As was mentioned before, the FS segments which give the major contributions to the formation of the transverse Fermi-liquid mode are nearly paraboloidal in shape, therefore the results of the work \cite{19} may be used to qualitatively estimate the significance of the surface scattering of  electrons under the conditions of the anomalous skin effect.  We assume for simplicity that the diffuse scattering is characterized by a constant $P\ (0<P<1).$ When $ P=0, $ the reflection of electrons is purely specular, whereas $P=1$ corresponds to the completely diffuse reflection. Adopting the expression (2) to describe electrons spectrum one could obtain:
\bea 
 \sigma_{n}^\pm & =& \pm \frac{iNe^2}{3m_\perp \omega} \chi_\pm \left(\frac{1}{\theta_n^\pm} + \frac{2(\chi_\pm^*/\chi_\pm)^2}{\theta_n^{*\pm}} \right),
\\ \nn\\
\sigma_{nn'}^\pm &=& \frac{4}{3}\frac{Ne^2}{m_\perp \omega} \lambda \frac{v_0}{\omega L} \chi_\pm^2 \left(\frac{1}{1\mp \lambda s_\pm} \frac{1}{\theta_n^\pm \theta_{n'}^\pm} \right.
\nn\\  && +         \left.
\frac{2(\chi_\pm^*/\chi_\pm)^4}{1 \mp \lambda s_\pm^* \chi_\pm^*/\chi_\pm} \frac{1}{\theta_n^{*\pm} \theta_{n'}^{*\pm}} \right)
   \eea
 where $N$ is the electrons density,
     \be   
 \begin{array}{l}
 s_\pm = i \tan (L\Omega \chi_\pm / v_0), \quad
 s_\pm^* = i \tan (L \Omega \chi_\pm^* /v_0),  \\
\theta_n^\pm = \chi_\pm^2 -q_n^2 ,  \\
 \theta_n^{*\pm} =\chi_\pm^{*2} - q_n^2 \equiv \chi_\pm^2 \mp \alpha_2
\chi_\pm - q_n^2  . \\
 \end{array} 
 \ee
 The parameter $\lambda = P/(2 - P) $ characterizes the strength of the diffuse component in the electron scattering from the surfaces of the metal film.

Comparing the expressions (32) and (33) we conclude that $ \sigma_n^\pm $ predominates over $ \sigma_{nn'}$ in magnitude when $ \lambda \xi \delta/L \ll \omega/\Omega.$ Assuming that the anomaly parameter $ \xi \sim 10^2, $ the skin depth $\delta \sim 10^{-6} $m, and $\omega \sim \Omega $ we conclude that the roughness of the surface does not affect the transmission coefficient if the film thickness $L $ is not smaller than $10^{-4}$m. For thinner films the surface roughness may bring noticeable changes into the transmission. For instance, when $L \sim l \sim 10^{-5} $m, we may neglect the diffuse contribution to the electrons reflection at the surfaces of the film when $ \lambda < 0.1\ (P< 0.2).$ In further calculations we assume the film surfaces to be smooth enough, so that we could treat the electrons reflection from the metal film surfaces as nearly specular. Correspondingly, we omit the second term in the expression (31). 

Substituting the resulting expressions for $j_n^\pm $ into Eq. (29) we get:
   \be 
  E_n^\pm = \mp \frac{\omega}{c} F_\pm (\omega,q_n) [(-1)^n b_\pm (L) - b_\pm (0)].
  \ee
  Here, we introduced the notation:
   \be 
  F_\pm (\omega,q_n) = \left (q_n^2 - \frac{4\pi i \omega}{c^2} \sigma_n^\pm\right)^{-1}.
  \ee
  Now, using these expressions for the Fourier transforms we get the relations for the electric and magnetic fields at the interfaces $ z =0 $ and $ z = L:$
   \bea 
  E_\pm (0) = \frac{c}{4\pi} \left[ Z_\pm^{(0)} b_\pm (0) - Z_\pm^{(1)} b_\pm (L) \right],
  \\\nn\\
  E_\pm (0) = \frac{c}{4\pi} \left[ Z_\pm^{(1)} b_\pm (0) - Z_\pm^{(0)} b_\pm (L) \right],
  \eea
  where the surface impedances are given by:
   \bea 
  Z_\pm^{(0)}\!&=&\! \pm \frac{8\pi\omega}{Lc^2} \sum_{n=0} \left(1- \frac{1}{2} \delta_{n0} \right) F_\pm (\omega,q_n),
  \\\nn\\ 
  Z_\pm^{(1)}\! &=& \!\pm \frac{8\pi\omega}{Lc^2} \sum_{n=0} \left(1- \frac{1}{2} \delta_{n0} \right)(-1)^n F_\pm (\omega,q_n).
   \eea
  To get the expression for the transmission coefficient which is determined by the ratio of the amplitudes of the transmitted field $(E_t)$ at $ z=L $ and the incident field $(E_i)$ at $ z=0$ we use the Maxwell boundary conditions:
   \be 
 2E_i^\pm = E_\pm (0) + b_\pm (0), \qquad E_t^\pm = b_\pm (L).
  \ee
  Then we define $ T_\pm = \big|E_t^\pm / E_i^\pm \big| $ where
$ E_t^\pm/ E_i^\pm = [E_\pm (L) + b_\pm (L)]/[E_\pm (0) + b_\pm(0)] .$ Assuming that the transmission is small $(T_\pm \ll 1) $ we get the asymptotic expression:
   \be 
  \frac{E_t^\pm}{E_i^\pm} \approx \frac{c}{4\pi} Z_\pm^{(1)}
  \ee
  where $ Z_\pm^{(1)} $ is given by the  relation (40).

  Therefore, keeping the $``-"$ polarization  we can start from the following expression for the transmission coefficient: 
       \be 
T = \frac{4 i\omega}{L c} \sum_{n=0} (-1)^{n+1}
\left( 1 - \frac{1}{2} \delta_{n0} \right) F_-(\omega,q_n),
                \ee
   Using the Poisson's summation formula:
          \be 
\sum_{n=0} y (q_n) =
\sum_{r=-\infty}^\infty \int_0^\infty
y \left (\frac{\pi}{L}x \right) \exp (2 \pi i r x) dx,
                       \ee
 we convert the expression for the transmission coefficient to the form:
          \be 
T = \frac{2}{\pi} \frac{\omega}{c} \int_{-\infty}^\infty
\mbox {sign} (q) \mbox{cosec} (Lq) F_- (\omega,q) dq.
                         \ee
  where $\mbox{sign}(q) $ it the sign function: $\mbox{sign}(q) = |q|/q.$   An important contribution to the integral (45) comes from the poles of the function $ F_-(\omega,q), $ i.e. the roots of the dispersion equation (16) for the relevant polarization.
  The contribution from the considered low frequency mode to the transmission coefficient is equal to a residue from the appropriate pole of the integrand in the expression (45).

 When $ dA/dp_z $ gets its extremal values at the inflection lines $ (p_z = \pm p^*)$ the contribution $T_1 $ from this wave to the transmission coefficient is:
   \bea 
  T_1 \!&\approx&\! \frac{\rho_s}{\xi} \frac{v_0}{c} \alpha_2^2 \tilde \omega^3 (|\alpha_2|\tilde \omega)^{-7s/2}  (1-\tilde\omega)^{-3/(2s-5)}
  \nn \\  \!\!\!&&\!  \times
\left[\sin^2 \left(\frac{L}{l} \Omega\tau (1-\tilde\omega)\right) + \sinh^2 \left(\frac{L}{l}\right)\right]^{-1/2}
  \eea
  where  $\tilde \omega = \omega/\Omega(p^*),$ and $ \rho_s $ is the dimensionless factor of the order of unity:
   \be 
  \rho_s = \frac{2s-3}{2s -5}\, \frac{Q_0}{Q_2} (x^*)^2 (\nu_s)^{(2s-3)/(2s-5)}
   \ee
  The size oscillations of the transmission coefficient arising due to the low frequency cyclotron wave could be observed in thin films whose thickness is smaller than the electrons mean free path $(L \ll  l).$
 Under this condition we can obtain the following
estimates for $ |T_1| $ in a typical metal in a magnetic field of the order of $ 5T$, and for the shape parameter $ s=3:$ 
          \be 
T_1 \sim \left(10^{-10}\div 10^{-11} \right) l/L. 
\ee
  Size oscillations of the transmission coefficient described by the expression (43) are shown in the figures 4,5.  When $(s=3)$ (see Fig. 4) the oscillations amplitudes accept values $\sim 10^{-8} \div 10^{-9} $ depending on the ratio $ L/ l. $ The values of such order can be measured in experiments on the transmission of electromagnetic waves through thin metal films. However, the oscillations magnitudes may reach significantly greater values when the shape parameter increases. As displayed in the Fig. 5, $ T_1 $ can reach the values of the order of $ 10^{-6} $ when $s = 5. $

\begin{figure}[t]
\begin{center}
\includegraphics[width=7.5cm,height=7.3cm]{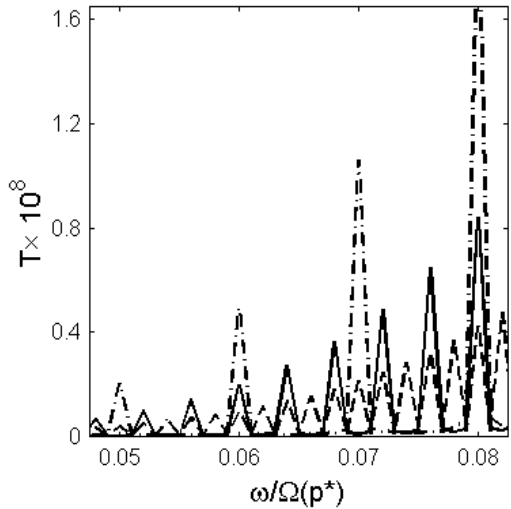}
\caption{ Size oscillations in the transmission coefficient for the transverse electromagnetic wave traveling through a metal film which originate from the low frequency Fermi-liquid mode. The curves are plotted at $ \alpha_2 =-0.2,\ s = 3.\ \Omega\tau \sim 50, \xi = 10^3, \ L/l = 0.01 $ (dash-dot line); $0.025 $ (solid line) and $ 0.05 $ (dashed line).}
 \label{rateI}
\end{center}
\end{figure}

Under considered conditions the transmission coefficient also includes a contribution $T_2 $ from electrons corresponding to the vicinities of those cross-sections of the Fermi surface where the longitudinal component of their velocity becomes zero.  This contribution always exists under the anomalous skin effect. 
The most favorable conditions for observation of the size oscillations arising due to the Fermi-liquid wave in experiments are provided when $T_1 >T_2.$ It happens  when $ L \omega \xi > v_0.$ When the FS everywhere has a finite nonzero curvature the expression for $T_{2} $ can be written as follows \cite{20}: 
          \be 
T_2 \approx \frac{4}{3} \frac{v_0}{c} \frac{1}{\xi} \exp
\left ( - \frac{L \omega\tau \xi}{l} \right );
                      \ee
 In  magnetic fields $ \sim 5 T $ and for $ L \omega \sim v_0$ the contribution $T_2$ has the order of $ 10^{-10} \div 10^{-11}, $ i.e. the predominance of the term $T_1 $ over $T_2$ can be reached. Besides the contributions from the poles of $F_- (\omega,q) $ the transmission coefficient (45) includes a term $T_3$ originating from the branch points of this function in the $ q,\omega $ complex plane.
  These points cause
the Gantmakher--Kaner size oscillations of the transmission coefficient \cite{21}. However, for $ L \Omega > v_0, $ these oscillations have  a magnitude of the order of $ 10^{-9}\div 10^{-10} $ or less. So, the present estimates give grounds to expect that the size oscillations in the transmission coefficient of the electromagnetic wave through a thin film of a clean metal may include a rather significant, or even predominating contribution, which arises due to the low frequency $ (\omega \ll \Omega)$ Fermi-liquid mode. 

\begin{figure}[t]
\begin{center}
\includegraphics[width=7.5cm,height=7.3cm]{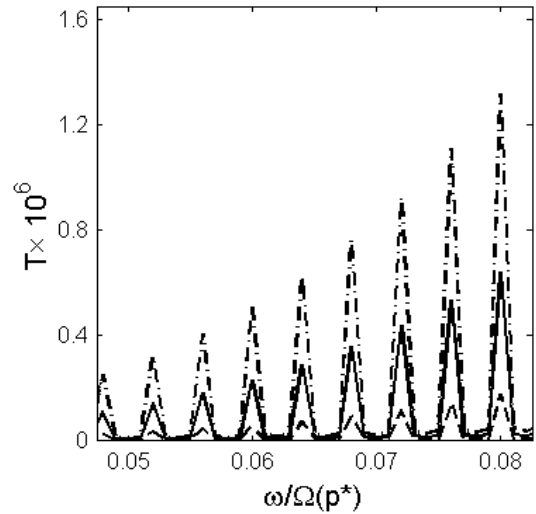}
\caption{ The dependence of the transmission from the FS shape near the inflection line. The curves are plotted for $ s=3 $ (dashed line), 4 (solid line) and 5 (dash-dot line), $ L/l = 0.025. $ The remaining parameters coincide with those used to plot the curves in the figure 4.}
 \label{rateI}
\end{center}
\end{figure}

Fermi surfaces of real metals are very complex in shape and most of them have inflection lines, so there are grounds to expect the low frequency Fermi-liquid waves to appear in some metals. Especially promising are such metals as cadmium, tungsten and molybdenium where collective excitations near the Doppler-shifted cyclotron resonance (dopplerons) occur \cite{12,13,14}. Another kind of interesting substances are quasi-two-dimensional conductors. Applying the external magnetic field along the FS axis and using the tight-binding approximation for the charge carriers, we see that the maximum longitudinal velocity of the latter is reached at the FS inflection lines where $ d^2 A/dp_z^2 = 0. $ So, we may expect the low frequency Fermi-liquid wave to appear at some of these substances along with the usual Fermi-liquid cyclotron wave.

\section{iv. conclusion}

It is  a common knowledge that electron-electron correlations in the system of conduction electrons of a metal may cause  occurences of some collective excitations (Fermi-liquid modes), whose frequencies are rather close to the cyclotron frequency at strong magnetic fields $(\Omega\tau \gg 1)$. Here we show that a Fermi-liquid wave can appear in clean metals at significantly lower frequencies $ (\tau^{-1} \ll \omega\ll\Omega).$ The major part in the wave formation is taken by the electrons (or holes) which move along the applied magnetic field with the maximum velocity $ v_0. $ Usually, such electrons belong to the vicinities of limiting points or inflection lines on the FS. When the FS possesses  nearly paraboloidal segments including these points/lines, the longitudinal velocity of the charge carriers slowly varies over such FS segments remaining close to its maximum value $ v_0.$ This strengthens the response of these ``efficient" electrons to the external disturbances. As a result the spectrum of the Fermi-liqud cyclotron wave may be significantly changed.  These changes were analyzed in some earlier works (see e.g. Ref. \cite{15}) assuming that the cyclotron mass of the charge carriers remains the same all over the FS. Under this assumption it was shown that the appropriate FS geometry at the segments where the maximum longitudinal velocity of electrons/holes is reached may cause the dispersion curve of the transverse Fermi-liquid cyclotron wave to be extended to the region of comparatively low frequencies $ (\omega\ll \Omega).$

In the present work we take into account the dependence of cyclotron mass of $ p_z. $ This more realistic analysis leads to the conclusion that one hardly may expect the above extension of the Fermi-liquid cyclotron wave spectrum in real metals. However, when the FS has the suitable geometry at the segments where the charge carriers with maximum longitudinal velocity are concentrated, the low frequency Fermi-liquid mode may occur in the metal alongside the usual Fermi-liquid cyclotron wave.
 This mode may cause a special kind of size oscillations in the transmission coefficient for an electromagnetic wave of the corresponding frequency and polarization incident on a thin metal film.

\section{Acknowledgments}

The author thanks G. M. Zimbovsky for help with the manuscript. This work was supported  by NSF Advance program SBE-0123654, DoD grant W911NF-06-1-0519,  and PR Space Grant NGTS/40091.


\begin{thebibliography}{99}
\bibitem{8} At $ q > q_m$ the wave is damped by the conduction electrons.
\end{thebibliography}
\end{document}